*Original Article*

# Analysis of Factors Affecting the Entry of Foreign Direct Investment into Indonesia (Case Study of Three Industrial Sectors in Indonesia)

**[1]Tracy Patricia Nindry Abigail Rolnmuch, [2]Yuhana Astuti**
[1,2]*School of Economics and Business, Telkom University, Bandung, Indonesia.*

***Abstract:*** *The realization of FDI and DDI from January to December 2022 reached Rp1,207.2 trillion. The largest FDI investment realization by sector was led by the Basic Metal, Metal Goods, Non-Machinery, and Equipment Industry sector, followed by the Mining sector and the Electricity, Gas, and Water sector. The uneven amount of FDI investment realization in each industry and the impact of the COVID-19 pandemic in Indonesia are the main issues addressed in this study. This study aims to identify the factors that influence the entry of FDI into industries in Indonesia and measure the extent of these factors' influence on the entry of FDI. In this study, classical assumption tests and hypothesis tests are conducted to investigate whether the research model is robust enough to provide strategic options nationally. Moreover, this study uses the ordinary least squares (OLS) method. The results show that the electricity factor does not influence FDI inflows in the three industries. The Human Development Index (HDI) factor has a significant negative effect on FDI in the Mining Industry and a significant positive effect on FDI in the Basic Metal, Metal Goods, Non-Machinery, and Equipment Industries. However, HDI does not influence FDI in the Electricity, Gas, and Water Industries in Indonesia.*

## I. INTRODUCTION

Indonesia has abundant natural resources to a large population, making it a great potential as a promising market for foreign investors [1]. The natural resources owned by Indonesia have the potential to attract some investors from local to foreign countries with the intention of developing their business [2], [3]. A stable and evenly distributed electricity supply is one of the important pillars of industrial growth and development in Indonesia. There has been an increase in electricity consumption in the PLN (*Perusahaan Listrik Negara*) network in Indonesia in the last 10 years, from 174 TWh in 2012 to 255 TWh in 2021, showing that electricity has the highest consumption growth compared to other types of energy [4]. Currently, fossil fuel power plants dominate national electricity production with a composition of between 66% and 80% [5].

Significant social progress is demonstrated by the increase in Indonesia's Human Development Index (HDI), which can be seen from the escalation in the dimensions of education, health, and public welfare [6]. The HDI in Indonesia has consistently experienced a high surge [6], but in the last four years, there has been a slowing trend in the HDI, especially from 2019 to 2020, where there were restrictions on activities during the wider spread of COVID-19. Along with the pandemic, HDI has improved and started to grow again in 2021 at 72.29 points.

The realization of Foreign Direct Investment (FDI) and Domestic Direct Investment (DDI) in the period January to December 2022 reached Rp1,207.2 trillion [7]. Indonesia has managed to attract significant foreign investment in recent years, particularly in the metals, mining, and energy (electricity, gas, and water) industries. The largest FDI realization by sector was led by the Mining sector, followed by the Basic Metals, Metal Goods, Non-Machinery, and Equipment sector, and the Electricity, Gas, and Water sector. The amount of FDI revenues from sectors in Indonesia has not been evenly distributed, which indicates the need for focused policies and strategies to support industries that are not fully developed to create a balance of Indonesia's economic growth across industries. The Mining Industry is the industry with the highest FDI revenues compared to other sectors from 2012-2021. This sector recorded more than US$35 billion in FDI revenues. The second and third sectors that receive the largest FDI are the Basic Metal, Metal Goods, Non-Machinery and Equipment Industry and the Electricity, Gas, and Water Industry. Certain regions still dominate the current distribution of FDI in Indonesia. In the January-September period of 2023, Java Island dominated the FDI realization with an inflow value of US$263,279.6 million [8]. Java is followed by Sulawesi, Sumatra, Maluku and Papua, Kalimantan, then Bali and Nusa Tenggara.

The Mining Industry in Indonesia faces a multitude of complex challenges that impact the sustainability and growth of the sector. Environmental impacts not only damage ecosystems but also provoke resistance from local communities directly affected by mining activities. Commodity price fluctuations pose economic challenges for the mining sector, influencing government investment policies. The Indonesia Investment Coordinating Board strives to create a stable investment climate and attract investors by offering fiscal and non-fiscal incentives for mining investments [9]. Meanwhile, the Basic Metal, Metal Goods, Non-Machinery, and Equipment Industries serve as the foundation for the manufacturing and construction sectors, with basic metal products being essential raw materials for various industries, including automotive, electronics, and household appliances. A primary challenge faced by Indonesia's basic metal industry is the volatility of global raw material prices. Prices for raw materials such as metal ores and minerals often fluctuate significantly due to external factors such as global market conditions, international trade policies, and changes in demand from major countries. This price instability can disrupt long-term financial planning and investment, as well as hinder innovation and the development of new products. The Electricity, Gas, and Water Industry is essential in daily life and significantly impacts the national economy. Infrastructure projects, such as the construction of gas and water-based power plants and the development of distribution networks, are major attractions for investors. Foreign investment significantly contributes to large projects like power plants and gas infrastructure. For instance, in 2021, foreign investment in the electricity sector exceeded US$2,938,583 thousand, encompassing various projects in steam power, gas, and renewable energy. Despite government efforts to create a supportive investment environment, investors still face regulatory hurdles. Complicated and time-consuming permitting processes are major obstacles, and policy instability and a lack of coordination among government agencies can create uncertainty for investors.

Based on the explanation in the background above, changes in mining commodity prices, fluctuations in metal materials, and regulatory barriers are some of the main problems in the three industries previously mentioned. Then, the uneven amount of FDI realization in each industry and the COVID-19 pandemic in Indonesia have made FDI realizations inconsistent. On the other hand, foreign investors' interest is shaped by the availability of electricity infrastructure and the caliber of the workforce. A key factor is the government's strategy to entice FDI by ensuring that electrical energy is readily available to both the public and the industrial sector. Skilled and trained human resources are one of the key factors in determining the success of a country. Similarly, FDI is affected by robust economic activity and sufficient infrastructure [10]. The explanation that has been described previously gets the intend which is to find out and determine how much influence from the factors affect the entry of FDI in the Mining Industry, Basic Metal Industry, Metal Goods, Non-Machinery, and Equipment Industry, and Electricity, Gas and Water Industry in Indonesia.

## II. LITERATURE REVIEW

Foreign direct investment (FDI) is a crucial means for a country to achieve development and has become the main foreign funding source for many developing nations, outpacing government funding for development, private loans, portfolio equity, and remittances [11]. Indonesia has the potential for high FDI inflows; one of the driving factors is the availability of natural resources. The availability of natural resources makes Indonesia superior in the wealth of production factors so that it can attract investors to acquire various manufacturing businesses in Indonesia [10]. FDI fosters economic growth by enabling capital formation, technology transfer, and enhanced productivity. This relates to the assets created by investors to run foreign companies, including establishing ownership and controlling interests in those companies [12]. This explanation can be drawn that FDI is a form of direct investment by domestic companies in foreign countries that includes purchasing shares establishing and operating factories. FDI often brings innovation, such as new technology, management, and more efficient business practices, into the country [13]. Although some of the benefits of this FDI flow back to foreign investors, with this investment, the economy's capital stock can increase, which means more wages and productivity [14]. With direct investments from foreign companies into the host country, including establishing factories for products such as computers, smartphones, or software like applications and operating systems, these companies will increase imports of information technology-based goods in large quantities to meet their production needs [15].

Electric power is a type of secondary energy that is generated, transmitted, and distributed for various purposes. However, electricity used for communication, electronics, or signaling is not included. Electric power must always be improved to spread evenly due to its crucial and strategic role in achieving national development objectives. This nexus between economic growth and electricity suggests that as economic activities requiring electricity increase, the construction of additional power plants will also rise to meet the expanding energy demand [16]. Electrical energy significantly contributes to the global energy reserve, playing a key role in achieving sustainable economic growth and development [17]. A study by Nepal and Paija stated that a shortage in the supply of electrical energy can indirectly hinder future economic growth [18].

The Human Development Index (HDI) is a benchmark introduced by the United Nations Development Programme (UNDP) to measure human development (HD) from a multidimensional perspective [19]. HDI is a composite measure that evaluates the average achievement across three fundamental dimensions of human development: health and longevity, knowledge, and a decent standard of living [20]. HDI was introduced in the first Human Development Report (HDR) in 1990. UNDP uses four elements in human development, namely productivity, equity, sustainability, and empowerment. The calculation of HDI involves several calculation components, including the life expectancy of babies at birth, estimated years of schooling, average years of schooling, and gross national income (GNI) per capita. Life expectancy from birth serves as an indicator of lifespan and healthy living conditions. The average years of schooling and expected years of schooling indicators reflect the duration of education for individuals aged 25 and older. The third indicator of GNP per capita reflects the per capita expenditure of a country [10]. A study from Indrajaya and Iskamto states that HDI and economic growth significantly and negatively affect poverty reduction in Indonesia [21].

This research uses four variables. The independent variables are electricity and HDI, while the dependent variable is FDI. This research has a period from 2012 to 2021. There are differences in each result because, in 2016, there was an Economic Census, which resulted in missing data on the electricity variable, so interpolation was carried out on the missing data. The author also realizes that from 2012-2021, there was a COVID-19 pandemic crisis. To explore in further depth, this study intends to find the relationship between important factors such as electricity and HDI with FDI in three industries in Indonesia. Then, the data in this study are sourced from the Indonesia Central Statistics Agency and the Indonesia Investment Coordinating Board. Based on research conducted by Budiono and Purba, the variables of electricity, clean water, and HDI exert a highly positive influence on FDI. The study also reveals a strong correlation between each Indonesian province and the regression of electricity, water, and HDI. However, in this study, only electricity and HDI are examined on FDI because electricity and water are similar in categorizing as public infrastructure. The framework and the hypothesis of this research are as follows:

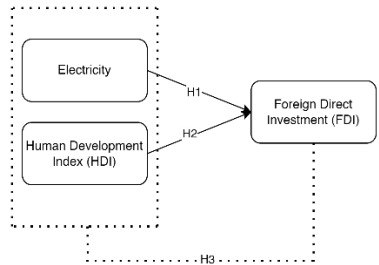

**Fig. 1 Research Framework**

Hypothesis 1: Electricity has a significant effect on FDI
Hypothesis 2: HDI has a significant effect on FDI
Hypothesis 3: Independent variables have a significant effect on FDI

In analyzing the data, this study involves stages of econometric testing using time-series data to determine the appropriate model. Moreover, this study also uses data analysis methods and techniques by conducting stationarity tests, correlation matrices, F tests, t-tests, coefficient of determination tests, and classical assumption tests. In addition, this study uses the Ordinary Least Square (OLS) model or multiple linear regression analysis, which refers to the model below [22].

$$FDI_{MIN} = \alpha + \beta_1 \text{ELCT} + \beta_2 \text{HDI} + \varepsilon \quad (1)$$

$$FDI_{BM} = \alpha + \beta_1 \text{ELCT} + \beta_2 \text{HDI} + \varepsilon \quad (2)$$

$$FDI_{EGW} = \alpha + \beta_1 \text{ELCT} + \beta_2 \text{HDI} + \varepsilon \quad (3)$$

where: $FDI_{MIN}$: FDI for Mining Industry
$FDI_{BM}$: FDI for Basic Metal, Metal Goods, Non-Machinery, and Equipment Industry
$FDI_{EGW}$: FDI for Electricity, Gas, and Water Industry
α: C
β: Coefficient
ε: *Error*
ELCT: Electricity as an independent variable
HDI: HDI as independent variable

## III. RESULTS AND DISCUSSION

*A) Mining Industry*

**Table 1: Stationarity Test for FDI$_{MIN}$**

| Variables on Lag=1 | Augmented Dickey-Fuller test statistic ($2^{nd}$ difference) |
|---|---|
| FDI$_{MIN}$ | 0.0408 |
| ELCT | 0.0142 |
| HDI | 0.0113 |

The results from Table 1 show that the unit root test in $2^{nd}$ difference is stationary for all variables. This is considered to qualify for the stationary test.

**Table 2: Correlation Matrix for FDI$_{MIN}$**

| | FDI$_{MIN}$ | ELCT | HDI |
|---|---|---|---|
| **FDI$_{MIN}$** | 1 | 0.023333 | -0.676908 |
| **ELCT** | 0.023333 | 1 | 0.438859 |
| **HDI** | -0.676908 | 0.438859 | 1 |

The results show that the correlation between FDIMIN and electricity, with a coefficient of 0.23333, is very weak and positive, with almost no linear relationship between the variables. The correlation between FDIMIN and HDI with a coefficient of -0.676908 is strongly negative, indicating that when FDIMIN increases, HDI will decrease significantly. Then, the correlation between electricity and HDI with a coefficient of 0.438859 has a moderate positive correlation, indicating that there is a positive relationship between the variables.

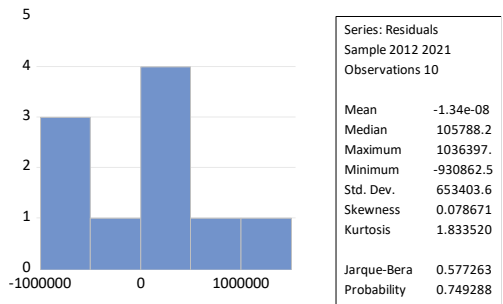

**Fig. 2 Normality Test for FDI$_{MIN}$**

The test results in Figure 2 show a probability value of 0.749288. This value is more than 5%, meaning that the data for the FDIMIN variable is normally distributed.

**Table 3: Multicollinearity Test for FDI$_{MIN}$**

| Variable | Coefficient Variance | Uncentered VIF | Centered VIF |
|---|---|---|---|
| C | 1.75E+15 | 31913.20 | NA |
| ELCT | 305.5454 | 26.61540 | 1.238539 |
| HDI | 3.56E+11 | 32687.45 | 1.238539 |

In Table 3, the Centered VIF on Electricity is 1.238539, and the HDI is 1.238539, which means the VIF value is less than 10, so it qualifies for the multicollinearity test.

**Table 4: Heteroscedasticity Test for FDI$_{MIN}$**

| F-statistic | 4.779445 | Prob. F(2,7) | 0.0491 |
| Obs*R-squared | 5.772663 | Prob. Chi-Square(2) | 0.0558 |
| Scaled explained SS | 1.178850 | Prob. Chi-Square(2) | 0.5546 |

The results in Table 4 above use the Breusch-Pagan-Godfrey heteroscedasticity test on the FDIMIN variable that shows the prob. value on Obs*R-squared is at 0.0558, so there is no heteroscedasticity.

**Table 5: Autocorrelation Test for FDI$_{MIN}$**

| Mean dependent var | 3598941. |
|---|---|
| S.D. dependent var | 1014705 |
| Akaike info criterion | 30.11242 |
| Schwarz criterion | 30.20319 |

| Hannan-Quinn criter. | 30.01284 |
|---|---|
| Durbin-Watson stat | 1.592468 |

The autocorrelation test results in Table 5 show the Durbin-Watson (DW) value at 1.5924, which means there is no autocorrelation.

**Table 6: Multiple Linear Regression Analysis for $FDI_{MIN}$**

| Variable | Coefficient | Std. Error | t-Statistic | Prob. |
|---|---|---|---|---|
| C | 13549606 | 41854190 | 3.224543 | 0.0146 |
| ELCT | 25.60898 | 17.47986 | 1.465056 | 0.1863 |
| HDI | -1875813. | 597004.4 | -3.142043 | 0.0163 |

The data processing results of the regression equation in Table 6 can be concluded that the coefficient of electricity is 25.60898. This indicates that if PLN produces and distributes electricity to customers, FDI in this industry will increase by US$25.60898 million. Then, the regression coefficient on HDI is -1875813, which means that if the value of the HDI index increases, then FDI decreases by US$1,875,813 million. So that the equation obtained is as follows:

$$FDI_{MIN} = 13549606 + 25.60898 \text{ ELCT} - 1875813 \text{ HDI} + \varepsilon \qquad (4)$$

B) *Basic Metal, Metal Goods, Non-Machinery, and Equipment Industry*

**Table 7: Stationarity Test for $FDI_{BM}$**

| Variables on Lag=1 | *Augmented Dickey-Fuller test statistic (1st difference)* |
|---|---|
| $FDI_{BM}$ | 0.1660* |
| ELCT | 0.0252 |
| HDI | 0.7407* |

The results from Table 7 show that in the unit root test in 1st difference, only the Electricity variable is stationary, while the FDIBM and HDI variables are not stationary, causing them not to qualify for the stationary test.

**Table 8: Correlation Matrix for $FDI_{BM}$**

|  | $FDI_{BM}$ | ELCT | HDI |
|---|---|---|---|
| $FDI_{BM}$ | 1 | 0.502757 | 0.803707 |
| ELCT | 0.502757 | 1 | 0.827729 |
| HDI | 0.803707 | 0.827729 | 1 |

The results show a correlation between FDIBM and electricity with a coefficient of 0.502757, which is medium positive, indicating a significant positive relationship between the variables. The correlation between FDIBM and HDI with a coefficient of 0.803707 is strongly positive, indicating that when FDIBM increases, HDI will also increase significantly. Then, the correlation between electricity and HDI with a coefficient of 0.827729 has a strong positive correlation, indicating a significant relationship between variables.

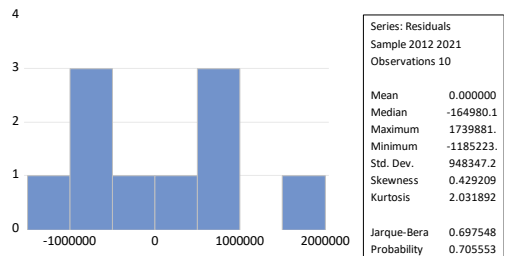

**Fig. 3 Normality Test for $FDI_{BM}$**

The test results in Figure 3 show a probability value of 0.705553. This value is more than 5%, meaning that the data for the FDIBM variable is normally distributed.

**Table 9: Multicollinearity Test for $FDI_{BM}$**

| Variable | Coefficient Variance | Uncentered VIF | Centered VIF |
|---|---|---|---|
| C | 8.74E+15 | 75543.76 | NA |
| ELCT | 7074.366 | 324.8309 | 3.175968 |
| HDI | 1.93E+11 | 83819.93 | 3.175968 |

In Table 9, the Centered VIF on Electricity is 3.175968, and the HDI is 3.175968, which means the VIF value is less than 10, so it qualifies for the multicollinearity test.

Table 10: Heteroscedasticity Test for $FDI_{BM}$

| | | | |
|---|---|---|---|
| F-statistic | 0.760322 | Prob. F(2,7) | 0.5026 |
| Obs*R-squared | 1.784649 | Prob. Chi-Square(2) | 0.4097 |
| Scaled explained SS | 0.451186 | Prob. Chi-Square(2) | 0.7980 |

The results in Table 10 above using the Breusch-Pagan-Godfrey heteroscedasticity test on the $FDI_{BM}$ variable show the prob. value on Obs*R-squared is at 0.4097, so there is no heteroscedasticity.

Table 11: Autocorrelation Test for $FDI_{BM}$

| | |
|---|---|
| Mean dependent var | 3255962. |
| S.D. dependent var | 1824430 |
| Akaike info criterion | 30.85747 |
| Schwarz criterion | 30.94824 |
| Hannan-Quinn criter. | 30.75789 |
| Durbin-Watson stat | 1.331006 |

The autocorrelation test results in Table 11 show the Durbin-Watson (DW) value at 1.331, which means there is no autocorrelation.

Table 12: Multiple Linear Regression Analysis for $FDI_{BM}$

| Variable | Coefficient | Std. Error | t-Statistic | Prob. |
|---|---|---|---|---|
| C | -33384608 | 93462820 | -3.571967 | 0.0091 |
| ELCT | -123.9731 | 84.10925 | -1.473953 | 0.1840 |
| HDI | 4877890. | 1387543 | 3.515487 | 0.0098 |

The data processing results of the regression equation in Table 12 can be concluded that the coefficient of electricity is -123.9731. This indicates that if PLN produces and distributes electricity to customers, FDI in this industry will decrease by US$123.9731 million. Then, the regression coefficient on HDI is 4877890, which means that if the value of the HDI index increases, then FDI increases by US$4,877,890 million. So that the equation obtained is as follows:

$$FDI_{BM} = -33384608 - 123.9731 \text{ ELCT} + 4877890 \text{ HDI} + \varepsilon \quad (5)$$

*C) Electricity, Gas, and Water Industry*

Table 13: Stationarity Test for $FDI_{EGW}$

| Variables on Lag=1 | Augmented Dickey-Fuller test statistic ($1^{st}$ difference) |
|---|---|
| $FDI_{EGW}$ | 0.0477 |
| ELCT | 0.0715* |
| HDI | 0.7407* |

The results from Table 13 show that in the unit root test in $1^{st}$ difference, only the Electricity variable is stationary, while the $FDI_{BM}$ and HDI variables are not stationary, causing them not to qualify for the stationary test.

Table 14: Correlation Matrix for $FDI_{EGW}$

| | $FDI_{EGW}$ | ELCT | HDI |
|---|---|---|---|
| $FDI_{EGW}$ | 1 | 0,736983 | 0,736852 |
| ELCT | 0,736983 | 1 | 0,757817 |
| HDI | 0,736852 | 0,757817 | 1 |

The results show a correlation between $FDI_{EGW}$ and electricity with a coefficient of 0.736983, which is strongly positive, indicating a significant positive relationship where if $FDI_{EGW}$ increases, then electricity tends to increase as well. The correlation between $FDI_{EGW}$ and HDI with a coefficient of 0.736852 is strongly positive, indicating that when $FDI_{EGW}$ increases, HDI will also increase significantly. Then, the correlation between electricity and HDI with a coefficient of 0.757817 has a strong positive correlation, indicating a significant relationship between the variables.

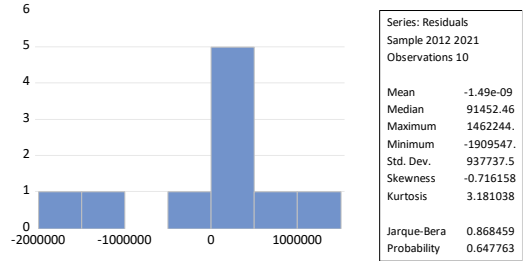

**Fig. 4 Normality Test for FDI$_{EGW}$**

The test results in Figure 4 show a probability value of 0.647763. This value is more than 5%, meaning that the data for the FDI$_{EGW}$ variable is normally distributed.

**Table 15: Multicollinearity Test for FDI$_{EGW}$**

| Variable | Coefficient Variance | Uncentered VIF | Centered VIF |
|---|---|---|---|
| C | 6.47E+15 | 57183.74 | NA |
| ELCT | 3968.731 | 176.7854 | 2.349002 |
| HDI | 1.39E+11 | 61994.70 | 2.349002 |

In Table 15, the Centered VIF on Electricity is 2.349002, and the HDI is 2.349002, which means the VIF value is less than 10, so it qualifies for the multicollinearity test.

**Table 16: Heteroscedasticity Test for FDI$_{EGW}$**

| F-statistic | 1.211682 | Prob. F(2,7) | 0.2322 |
|---|---|---|---|
| Obs*R-squared | 3.410750 | Prob. Chi-Square(2) | 0.1817 |
| Scaled explained SS | 1.822549 | Prob. Chi-Square(2) | 0.4020 |

The results in Table 16 above using the Breusch-Pagan-Godfrey heteroscedasticity test on the FDI$_{EGW}$ variable show that the prob. value on Obs*R-squared is at 0.1817, so there is no heteroscedasticity.

**Table 17: Autocorrelation Test for FDI$_{EGW}$**

| Mean dependent var | 3225265. |
|---|---|
| S.D. dependent var | 1516957. |
| Akaike info criterion | 30.83497 |
| Schwarz criterion | 30.92574 |
| Hannan-Quinn criter. | 30.73539 |
| Durbin-Watson stat | 1.769818 |

The autocorrelation test results in Table 17 show the Durbin-Watson (DW) value at 1.769, which means there is no autocorrelation.

**Table 18: Multiple Linear Regression Analysis for FDI$_{EGW}$**

| Variable | Coefficient | Std. Error | t-Statistic | Prob. |
|---|---|---|---|---|
| C | -99922875 | 80406242 | -1.242725 | 0.2540 |
| ELCT | 73.79919 | 62.99787 | 1.171455 | 0.2797 |
| HDI | 1380471. | 1179951 | 1.169939 | 0.2809 |

The data processing results of the regression equation in Table 18 can be concluded that the coefficient of electricity is 73.79919. This indicates that if PLN produces and distributes electricity to customers, FDI in this industry will increase by US$73.79919 million. Then, the regression coefficient on HDI is 1380471, which means that if the value of the HDI index increases, then FDI increases by US$1,380,471 million. So that the equation obtained is as follows:

$$FDI_{EGW} = -99922875 + 73.79919 \text{ ELCT} + 1380471 \text{ HDI} + \varepsilon \qquad (6)$$

*D) T-test, F-test, and Coefficient of Determination*
The results of the t-test on the three industries have been summarized in Table 19.

**Table 19: T-test for three industries in Indonesia**

|  | Mining Industry | Basic Metal, Metal Goods, Non-Machinery, and Equipment Industry | Electricity, Gas, and Water Industry |
|---|---|---|---|
| **ELCT** | 0.1863 | 0.1840 | 0.2797 |
| **HDI** | 0.0163 | 0.0098 | 0.2809 |
| **Hypothesis** | $H_1$ Rejected, $H_2$ Accepted | $H_1$ Rejected, $H_2$ Accepted | $H_1$ and $H_2$ Rejected |

Then, there are also the results of the F-test with the coefficient of determination test that have been carried out with the results in Table 20.

**Table 20: F-test and Coefficient of Determination for three industries in Indonesia**

|  | ELCT & HDI to FDI | Hypothesis ($H_3$) | Adjusted $R^2$ |
|---|---|---|---|
| **Mining Industry** | 0.045909 | Accepted | 53.4% |
| **Basic Metal, Metal Goods, Non-Machinery, and Equipment Industry** | 0.010254 | Accepted | 65.2% |
| **Electricity, Gas, and Water Industry** | 0.034495 | Accepted | 50.8% |

*E) Discussions*

While electricity is a critical component of mining operations, other factors are more important in attracting foreign investment to Indonesia's mining industry. One of them is the abundance of natural resource commodities, such as gold, tin, bauxite, copper, coal, nickel and silver [23]. The Ministry of Energy and Mineral Resources said that using of electricity is too much, and it is a challenge for the government to find ways to support industry with green energy. In terms of New Renewable Energy (EBT), in 2025, the government targets a 23% contribution of EBT to the electricity generation mix [24]. The entry of FDI into Indonesia through the establishment of manufacturing plants and companies in various sectors creates a large demand for electricity.

The significant negative influence of the human development index on foreign direct investment in Indonesia can be understood through the lens of economic cost structures. Higher HDI typically indicates better education, health, and living standards, which correlate with higher labor costs. Investors seeking cost-efficiency might find countries with lower HDI more attractive due to cheaper labor and operational costs. In Indonesia's context, as HDI improves, the cost of doing business increases, making it less appealing for foreign investors whose primary motive is to capitalize on lower production expenses. Thus, while a higher HDI represents socio-economic progress, it can inadvertently deter cost-sensitive FDI. This results in an adequate availability of skilled labor, becoming an attraction for foreign investors to invest in the industries [10]. These findings are corroborated by a study indicating that the HDI negatively and significantly impacts FDI [25]. However, some studies dispute this statement with their research stating that HDI with life expectancy has no influence on FDI supported by the significance value in the t test exceeding 5% [26], [27].

## IV. CONCLUSION

The outcomes of this research indicate that the electricity factor has no influence on FDI inflows in three industries: namely, the Mining Industry, the Basic Metal Industry, Metal Goods, the Non-Machinery and Equipment Industry, and the Electricity, Gas, and Water Industry. Then, the HDI factor has a significant negative effect on the Mining Industry and a significant positive effect in the Basic Metal, Metal Goods, Non-Machinery, and Equipment Industries, nevertheless not for the Electricity, Gas, and Water Industry, where HDI does not influence FDI in Indonesia.

The influence of electricity and HDI on FDI inflows in the Mining Industry is 46.6% and the remaining 53.4% is influenced by other independent variables not discussed in the context of this study. Then, the electricity and HDI factors that influence FDI inflows in the Basic Metal, Metal Goods, Non-Machinery, and Equipment Industry are 65.2%, and the remaining 34.8% is influenced by other independent variables not discussed in the context of this study. The electricity and HDI factors affecting FDI inflows in the Electricity, Gas, and Water Industry by 50.8% and the remaining 49.2% is influenced by other independent variables not discussed in the context of this study.

For better research, future researchers can investigate more deeply by using a variety of different industries and the number of samples with a longer data span than 10 years. Then, add other variables that can affect FDI besides electricity and HDI, such as gross domestic product (GDP), exchange rates, or inflation [28], to gain a deeper understanding of the factors that affect FDI inflows in Indonesia. The purpose of adding other variables and changing objects in future research is so that investors can understand more deeply the other factors that can affect the entry of FDI into a country, especially in Indonesia.

Also, it can find information about 20 other sectors recorded at the Indonesia Investment Coordinating Board. Ensuring widespread electricity provision and improving distribution are major policy concerns for both central and local governments to address the needs of industries and households throughout Indonesia. Furthermore, Indonesia is among the countries where the human development index—which measures living conditions, health, and education—is a top worldwide policy objective. Other factors or macroeconomics that influence the inflow of foreign investment into Indonesia can be taken into consideration for the largest recipient industries in Indonesia when playing a role in making policies related to foreign investment.

## V. REFERENCES

[1] Y. Astuti, "A Study on the Location Choice and Technology Spillovers of FDI: The Case of Indonesia's Manufacturing Industries," Waseda University, Tokyo, 2021. [Online]. Available: http://hdl.handle.net/2065/00081554
[2] I. A. Khan, H. Mokhlis, N. N. Mansor, H. A. Illias, L. Jamilatul Awalin, and L. Wang, "New trends and future directions in load frequency control and flexible power system: A comprehensive review," *Alexandria Engineering Journal*, vol. 71, pp. 263–308, May 2023, doi: 10.1016/j.aej.2023.03.040.
[3] A. N. Pramudhita and P. A. N. Mawangi, "SMART GRID UNTUK EFISIENSI KONSUMSI LISTRIK PADA PROSES PRODUKSI DI INDUSTRI MANUFAKTUR," *MATICS*, vol. 13, no. 1, pp. 7–12, Mar. 2021, doi: 10.18860/mat.v13i1.11566.
[4] Sekretariat Jenderal Dewan Energi Nasional, *Outlook Energi Indonesia 2022*. Jakarta: Sekretariat Jenderal Dewan Energi Nasional, 2022.
[5] Badan Pengkajian dan Penerapan Teknologi, *Outlook Energi Indonesia 2021: Perspektif Teknologi Energi Indonesia (Tenaga Surya untuk Penyediaan Energi Charging Station)*. Jakarta: Badan Pengkajian dan Penerapan Teknologi, 2021.
[6] Badan Pusat Statistik, "Indeks Pembangunan Manusia 2021," Jakarta, May 2022.
[7] Kementerian Investasi/BKPM, "Buku Statistik Realisasi Investasi Berdasarkan Sektor Tahun 2022," Jakarta, Jun. 2022.
[8] Kementerian Investasi/BKPM, "Perkembangan Realisasi Investasi: Triwulan III 2023," Jakarta, Oct. 2023.
[9] R. M. Fauzi and S. A. Nulhaqim, "Masalah Konflik Pertambangan di Indonesia," *Jurnal Kolaborasi Resolusi Konflik*, vol. 6, no. 1, pp. 34–41, 2024.
[10] S. Budiono and J. T. Purba, "Factors of foreign direct investment flows to Indonesia in the era of COVID-19 pandemic," *Heliyon*, vol. 9, no. 4, pp. 1–19, Apr. 2023, doi: 10.1016/j.heliyon.2023.e15429.
[11] UNCTAD, "World Investment Report 2023: Investing in Sustainable Energy for All," Geneva, 2023.
[12] S. Kusairi, Z. Y. Wong, R. Wahyuningtyas, and M. N. Sukemi, "Impact of digitalisation and foreign direct investment on economic growth: Learning from developed countries," *Journal of International Studies*, vol. 16, no. 1, pp. 98–111, Mar. 2023, doi: 10.14254/2071-8330.2023/16-1/7.
[13] D. Indrajaya and W. Driyastutik, "Human capital, unemployment, FDI, labor productivity and gross domestic product," *Jurnal Mantik*, vol. 8, no. 1, pp. 381–389, May 2024.
[14] N. G. Mankiw, *Principles of Economics*, 8th ed. New York: Cengage Learning, 2018.
[15] D. Indrajaya, D. U. Wardoyo, T. I. Santoso, and D. Iskamto, "Do Foreign Direct Investment and Innovation Influence ICT Goods Imports in ASEAN 6?," in *International Conference on Digital Business and Technology Management (ICONDBTM)*, Lombok: IEEE, Aug. 2023.
[16] R. F. Hirsh and J. G. Koomey, "Electricity Consumption and Economic Growth: A New Relationship with Significant Consequences?," *The Electricity Journal*, vol. 28, no. 9, pp. 72–84, Nov. 2015, doi: 10.1016/j.tej.2015.10.002.
[17] A. I. Lawal, I. Ozturk, I. O. Olanipekun, and A. J. Asaleye, "Examining the linkages between electricity consumption and economic growth in African economies," *Energy*, vol. 208, p. 118363, Oct. 2020, doi: 10.1016/j.energy.2020.118363.
[18] R. Nepal and N. Paija, "Energy security, electricity, population and economic growth: The case of a developing South Asian resource-rich economy," *Energy Policy*, vol. 132, pp. 771–781, Sep. 2019, doi: 10.1016/j.enpol.2019.05.054.
[19] B. K. Mangaraj and U. Aparajita, "Constructing a generalized model of the human development index," *Socioecon Plann Sci*, vol. 70, p. 100778, Jun. 2020, doi: 10.1016/j.seps.2019.100778.
[20] UNDP, "Human Development Report 2021-22: Uncertain Times, Unsettle Lives: Shaping our Future in a Transforming World," New York, 2022.
[21] D. Indrajaya and D. Iskamto, "Do Fund's Village, Economic Growth, Inequal Income Distribution, Unemployment Rate, and Human Development Index Affect Poverty in Indonesia?," in *Proceedings of the International Conference on Sustainable Collaboration in Business, Technology, Information, and Innovation*, Bandung: Springer, Nov. 2023, pp. 258–270. doi: 10.2991/978-94-6463-292-7_15.
[22] K. Lee, S. Im, and B. Lee, "Prediction of renewable energy hosting capacity using multiple linear regression in KEPCO system," *Energy Reports*, vol. 9, pp. 343–347, Nov. 2023, doi: 10.1016/j.egyr.2023.09.121.
[23] R. Gonzales, "Memaksimalkan potensi generasi muda di industri pertambangan untuk meningkatkan ekonomi Indonesia menuju Indonesia Emas 2045: Generasi Muda Untuk Bangsa," *Jurnal HIMASAPTA*, vol. 7, no. 1, pp. 39–50, Apr. 2022.
[24] Kementerian ESDM, *Rencana Strategis Direktorat Jenderal Energi Baru, Terbarukan dan Konservasi Energi 2020-2024*. Jakarta: Direktorat Jenderal Energi Baru, Terbarukan dan Konservasi Energi, 2021.
[25] Y. Astikawati and A. D. S. Sore, "Pengaruh Indeks Pembangunan Manusia dan Pertumbuhan Ekonomi Terhadap Investasi Asing di Indonesia," *Pacioli: Jurnal Kajian Akuntansi dan Keuangan*, vol. 1, no. 1, pp. 15–21, Jan. 2021.
[26] D. K. Syahrindra and D. Purnomo, "Determinan Penanaman Modal Asing di Indonesia Tahun 2012-2022," *Jurnal Menara Ekonomi : Penelitian dan Kajian Ilmiah Bidang Ekonomi*, vol. 10, no. 1, pp. 21–29, Apr. 2024, doi: 10.31869/me.v10i1.5277.
[27] D. Desmintari and L. Aryani, "Pengaruh Tata Kelola Pemerintahan, Indeks Pembangunan Manusia dan Total Productivity Terhadap Investasi Asing di Indonesia," *Jurnal Aplikasi Bisnis dan Manajemen*, vol. 8, no. 2, May 2022, doi: 10.17358/jabm.8.2.601.
[28] P. Bhujabal, N. Sethi, and P. C. Padhan, "Effect of institutional quality on FDI inflows in South Asian and Southeast Asian countries," *Heliyon*, vol. 10, no. 5, p. e27060, Mar. 2024, doi: 10.1016/j.heliyon.2024.e27060.